\documentclass[
 reprint,
 amsmath, amssymb,
 pre,
 aps,
 superscriptaddress
]{revtex4-2}

\usepackage{graphicx}
\usepackage{subfigure}
\usepackage{hyperref}
\usepackage{bm}
\usepackage{float}
\usepackage{xcolor}
\usepackage{amsmath}
\usepackage{braket} 

\newcommand\va[1]{\boldsymbol{#1}}
\newcommand\Expval[1]{\left\langle{#1}\right\rangle}
\newcommand\expval[1]{\braket{#1}}

\newcommand\Abs[1]{\left\vert{#1}\right\vert}
\definecolor{midnightblue}{RGB}{0, 0, 128}
\definecolor{mediumblue}{RGB}{0, 0, 205}
\definecolor{darkseagreen}{RGB}{66, 169, 87}
\definecolor{darkred}{RGB}{205,92,92}
\definecolor{deepskyblue}{HTML}{00BFFF}
\definecolor{goldenrod}{RGB}{218,165,32}  

\hypersetup{colorlinks=true, citecolor=blue, urlcolor=blue, linkcolor=blue}

\begin{document}

\preprint{APS/123-QED}

\title{Dynamical Tipping in a Quantum Limit Cycle}

\author{Ya-Xin Xiang}
\author{Zhengyang Bai}
     \email{zhybai@nju.edu.cn}
    \affiliation{National Laboratory of Solid State Microstructures and School of Physics,
	Collaborative Innovation Center of Advanced Microstructures, Nanjing University, Nanjing 210093, China}
	
\author{Yu-Qiang Ma}
     \email{myqiang@nju.edu.cn}
     \affiliation{National Laboratory of Solid State Microstructures and School of Physics,
	Collaborative Innovation Center of Advanced Microstructures, Nanjing University, Nanjing 210093, China}
	\affiliation{Hefei National Laboratory, Hefei 230088, China}

\begin{abstract}
Nonequilibrium systems maintain spatiotemporal order through energy dissipation and entropy production.
Here, we demonstrate this principle by engineering a quantum limit cycle that emerges from the interplay between a first-order absorbing-state phase transition and feedback mechanism coupling order and control parameters. 
The quantum limit cycle manifests dynamical correlations driven by multiplicative quantum noise and periodic traverses through tipping points, which act as a robust early-warning signal for state transitions.
This periodic crossing results in the enhanced dynamical susceptibility and transient rise of long-range correlations near each tipping event and their subsequent disappearance away from it. 
These critical transitions are accompanied by a sharp increase in the energy dissipation rate, leading to a stepwise accumulation of dynamical entropy production over time. 
Our results provide a way for realizing dynamical fluctuations and long-range correlations in strongly interacting driven-dissipative open quantum systems.
\end{abstract}
\maketitle

\section{Introduction}
Spontaneous oscillations, known as limit cycles (LCs), are exotic states unique to far-from-equilibrium systems. 
While extensively studied in classcial system, ranging from chemical and biological models \cite{lotka1920analytic,volterra1926fluctuations,prigogine1968symmetry,lefever1971chemical,
may1974biological,nisbet1976a,ertl1991oscillatory,epstein1996nonlinear,baras1996stochastic,
vanag2000oscillatory,boland2008how,boland2009limit}, they have recently been recognized for their profound resemblance to continuous time crystals in quantum systems~\cite{chan2015limit,tucker2018shattered,piazza2015self,chiacchio2019dissipation,barberena2019driven,
zhu2019dicke,buvca2019non,buca2019dissipation,kessler2019emergent,dogra2019dissipation,
kongkhambut2022obs,gao2023self, wu2024Dissipative,ding2024Ergodicity,jiao2025PhotoionizationInduced}.
In spatially extended systems, the emergence of periodic observables necessitates synchrony among distant constituents, typically achieved through the spontaneous breaking of both spatial and time-translation symmetries \cite{trager2021real, yang2023holo, giedirus2021six, dogra2019dissipation, bernien2017probing, smits2018obs}. Although the linear response to fluctuations in LCs has been thoroughly examined \cite{boland2008how, boland2009limit, chan2015limit, geysermans1997stochastic, Navarrete2017general, cabot2022quantum, carollo2022exact}, a fundamental question persists: Can non-equilibrium fluctuations and dynamical long-range correlations emerge in spatially extended systems in the absence of spatially periodic order?

\begin{figure}[b]
	\centering
	\includegraphics[width =8.6cm, keepaspectratio]{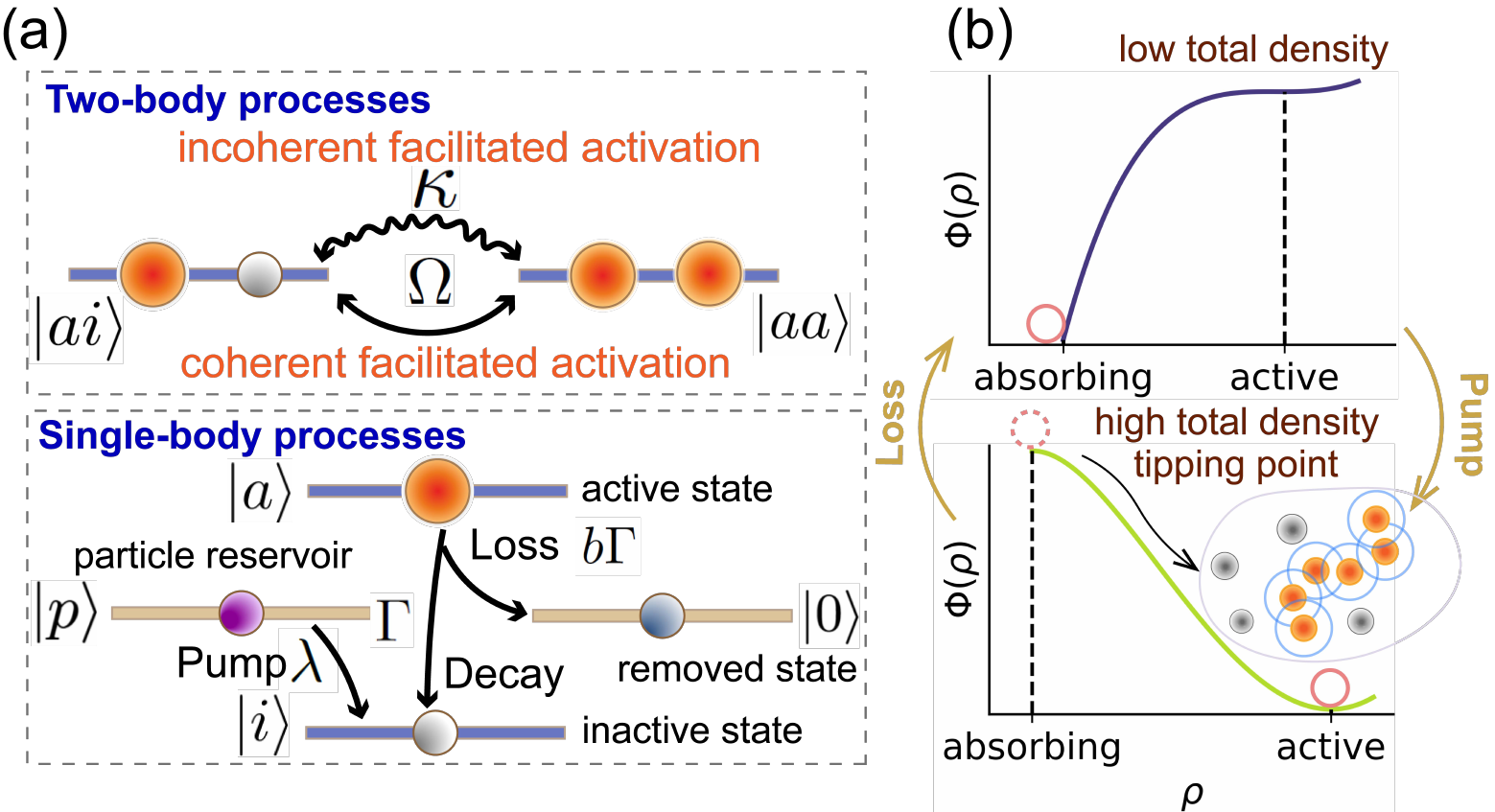}
	\caption{(a) Effective four-level scheme with timescale separation ($\Gamma,\kappa,\Omega\gg b\Gamma,\lambda$). 
    An inactive quantum emitter (state $\ket{i}$, gray sphere) can be activated via coherent ($\Omega$) and incoherent ($\kappa$) facilitated activation by a nearby emitter in the active state $\ket{a}$ (red sphere). Active emitters spontaneously decay either back to $\ket{i}$ or into a removed state $\ket{0}$ (blue sphere), causing emitter loss. Emitters in the pump state $\ket{p}$ (purple sphere) are incoherently transferred to the inactive state to compensate for losses.
    The system exhibits a first-order APT from a zero-excitation absorbing state to a finite-excitation active state as the total density increases. 
    This is captured by (b) the effective potential $\Phi(\rho)$ for active density $\rho$ at low (top) and high (bottom) total densities. The slow loss and pump processes couple the fast state-switching dynamics to the slow evolution of the total density, generating a limit-cycle state. 
    Within the cycle, a tipping point (dashed red circle) occurs when the absorbing state becomes linearly unstable, allowing a local excitation seed to proliferate throughout the system (insert; facilitation radius marked by a blue circle).}
	\label{fig:model}
\end{figure}

Here, we demonstrate such a many-body oscillatory state, intrinsically interwoven with long-range correlations, in an open quantum system. 
Our study is inspired by recent proposals to realize quantum contact processes and absorbing-state phase transitions (APTs) in cold Rydberg gases~\cite{marcuzzi2015non,marcuzzi2016absorbing,epidemic2017perez,gutierrez2017exp,roscher2018phe,
klocke2019control,helmrich2020signatures,gillman2021quantum,wintermantel2021epidemic,
brady2023MF,brady2024anomalous}. 
In this framework, the activation of an emitter is facilitated by its active neighbors~\cite{marcuzzi2016absorbing,helmrich2020signatures} [see Fig.~\ref{fig:model}(a)]. Within the quantum regime, the system exhibits bistability, where the order parameter (active population) undergoes a discontinuous jump at the critical threshold of the control parameter (total population).
Following the principles of self-organized bistability (SOB) \cite{buendia2020feedback,disanto2016sob}, we engineer a LC by cycling through the hysteresis loop.
This is realized by interfacing the fast transition dynamics with a slow feedback loop on the control parameter [Fig.~\ref{fig:model}(b)], as detailed in Ref. \cite{xiang2023self}.

The SOB-induced LC is characterized not only by periodic switching between absorbing and active phases but also by oscillatory long-range correlations. Although the LC remains linearly stable, it manifests a transiently diverging dynamical susceptibility. The transition from the absorbing to the active state is marked by a dramatic spike in this susceptibility, where a single activation seed proliferates catastrophically through the system—resembling a wildfire sweeping through a dense forest [see inset, Fig.~\ref{fig:model}(b)]. This phenomenon is deeply reminiscent of early-warning signals \cite{ding_tipping_2024,veraart2012recovery, van2007slow, scheffer2009early} observed near tipping points across diverse complex systems, ranging from ecosystems \cite{xu2023noneq, veraart2012recovery} and climate dynamics \cite{lenton2008tipping} to financial markets \cite{may2008complex}.

Furthermore, the broken time-translation symmetry is sustained by a continuous exchange of energy with the environment, which offsets the dynamical entropy production~\cite{wang2008potential, zhou2016construction}. Near the tipping point, each transition from the absorbing to the active state is accompanied by a sharp surge in dissipation and a burst of dynamical entropy. In contrast, the entropy production rate vanishes in the frozen absorbing state away from the threshold. Our work thus establishes a fundamental link between driven-dissipative quantum bistability and self-organized spatiotemporal order, highlighting the critical role of tipping points in far-from-equilibrium quantum systems.

\section{\label{sec:model}Model}
We consider a generic quantum contact model with $\ket{a}$ (active) and $\ket{i}$ (inactive) emitters~\cite{marcuzzi2016absorbing} [see Fig.~\ref{fig:model}(a)]. The active emitter can spontaneously become inactive (at rate $\Gamma$), and the inactive one can be activated only in the vicinity of active ones either coherently (at rate $\Omega$) or incoherently (at rate $\kappa$). 

Under the Markovian noise, the effective dynamics can be described by a Lindblad master equation for the density operator $\hat\rho$ ($\hbar = 1$),
\begin{equation}
\partial_t \hat\rho = -i[\hat H,\hat\rho] + \sum_{m} \mathcal{L}_{m}\hat\rho
\end{equation}
The coherent excitation is described by the Hamiltonian
\begin{equation}
	\hat H = \Omega \sum_l {\hat C}_l {\hat\sigma}_l^x,
\end{equation}
Here, $\Omega$ is the excitation rate, and the local operator ${\hat C}_l = \sum_{k\in\partial l}{{\hat\sigma}_k^{aa}}$ counts the number of active emitters in the neighborhood of emitter $l$, where $k,l$ label individual emitters and the summation $\sum_{k\in\partial l}$ runs over the effective nearest neighbors of the $l$-th emitter. The neighborhood is defined by a facilitation shell of radius $R_\text{fac}$ and thickness $r_\text{fac}$, such that two emitters are considered neighbors if their separation $R$ lies within the range $[R_\text{fac}-r_\text{fac},R_\text{fac}+r_\text{fac}]$~\cite{helmrich2020signatures}.

We define the local transition operators ${\hat\sigma}_l^{\alpha\beta}\equiv\ket{\alpha_l}\bra{\beta_l}$ ($\alpha, \beta= a, i, p, 0$)  and the Pauli operator ${\hat\sigma}_l^x=\hat\sigma^{-}_l+\hat\sigma^{+}_l$ and $\hat\sigma^{y}_l=i({\hat\sigma}_l^{-}-{\hat\sigma}_l^{+})$,. The latter serves to flip the quantum state via the ladder operators $\hat\sigma^{+}_l \equiv \hat\sigma_l^{ai}$ and $\hat\sigma^{-}_l \equiv \hat\sigma_l^{ia}$. 
The constraint operator ${\hat C}_l$ ensures that spin flips occur only in the presence of a neighboring active emitter, modeling a quantum analogue of stochastic branching or coagulation processes~\cite{marcuzzi2016absorbing,Haye2006nonequilibrium,xiang2023self}.

Local dissipative and incoherent processes are described by Lindblad terms of the form 
\begin{equation}
    \mathcal{L}_{m}\hat\rho = \sum_{l} \left[{\hat L}_{m,l}\hat\rho {\hat L}_{m,l}^\dagger -\frac{1}{2} \left\{{\hat L}_{m, l}^{\dagger}{\hat L}_{m, l},\hat\rho \right\}\right]
\end{equation}
where $m\in\set{d,p,e,a,b,c,s}$ labels each distinct process. 
The spontaneous decay of the active emitters is described by ${\hat L}_{d,l} = \sqrt{\Gamma}\hat\sigma^{-}_l$ with rate $\Gamma$, while dephasing of atomic coherences is described ${\hat L}_{p,l} = \sqrt{\gamma_\text{de}}{\hat\sigma}_l^{aa}$, with rate $\gamma_\text{de}$. 
We further introduce an removed state $\vert 0\rangle$, into which a fraction $b$ of active emitters decay via ${\hat L}_{e,l} = \sqrt{b\Gamma}\hat\sigma_l^{0a}$. 
To compensate for emitter loss,
an incoherent pump process $\ket{p}\to\ket{i}$ injects inactive emitters at a constant rate $\lambda$, implemented by ${\hat L}_{a,l}=\sqrt{\lambda}\hat\sigma_l^{ip}$. This can be realized by laser-coupling an auxiliary ground state to the $\ket{p}$ state to achieve population inversion~\cite{wang2000Gainassisted, Deng2007gain}.

Finally, the incoherent facilitated excitation and decay—corresponding to classical branching and coagulation processes—are incorporated via ${\hat L}_{b,l} = \sqrt{\kappa}{\hat C}_l{\hat\sigma}_l^{+}$ and ${\hat L}_{c,l} = \sqrt{\kappa}{\hat C}_l{\hat\sigma}_l^{-}$, respectively, both with rate $\kappa$.  
We also include a weak spontaneous activation at rate $\tau \ll \Gamma$ via $\hat{L}_{s,l} = \sqrt{\tau} \hat{\sigma}_l^+$ [see the supplementary material of Ref.~\cite{helmrich2020signatures} for a detailed derivation]. This process is only significant near the absorbing state—where the occupation of the active state is exactly zero—and leads to the additional activation term in Eq.~(\ref{eq:drhodt}) below.

To characterize the dynamical behavior of the system, we begin with the Heisenberg-Langevin equations of motion for the Pauli and density operators $\hat\sigma^{x/y/aa}_l$,
\begin{subequations}
	\begin{align}
		\partial_t \hat\sigma_l^{aa}  = & \tau\hat n_l - \Gamma\hat\sigma_l^{aa} + \Omega{\hat C}_l\hat\sigma_l^{y} + \kappa\hat C_l\left(\hat n_l - 2\hat\sigma_l^{aa}\right) + \hat\xi_l^{aa} \label{eq:EOM_qm_rr}\\
		\partial_t \hat\sigma_l^{x} = &-\frac{\kappa\hat N_l + \gamma}{2} \hat\sigma^x_l - \kappa\hat C_l\hat\sigma_l^{x} - \Omega{\hat P}_l\hat\sigma_l^{y} + \hat\xi_l^{x} \label{eq:EOM_qm_x}\\
		\partial_t \hat\sigma_l^{y} =&  -\frac{\kappa{\hat N}_l + \gamma}{2} \hat\sigma^y_l - \kappa{\hat C_l}\hat\sigma_l^{y} + \Omega{\hat P}_l\hat\sigma_l^{x} \label{eq:EOM_qm_y}\\
		&+ 2\Omega\hat C_l\left(\hat n_l - 2\hat\sigma_l^{aa}\right) + \hat\xi_l^{y}\notag\\
		\partial_t \hat n_l =& -b\Gamma\hat\sigma_l^{aa} + \lambda\hat\sigma_l^{pp}+ \hat\xi^{n}_l \label{eq:EOM_qm_n}
	\end{align}
\end{subequations}
where the local density $\hat n_l=\hat\sigma_l^{aa}+\hat\sigma_l^{ii}$, ${\hat N}_l =  \sum_{k \in\partial l} {{\hat n}_k} $, ${\hat P}_l = \sum_{k\in\partial l} {{\hat\sigma}_k^{x}}$, and $\gamma = \Gamma + \gamma_\text{de}$. 
The Langevin noise operators $\hat\xi_l^{x/y/aa/n}$ are determined by quantum fluctuation-dissipation relations \cite{xiang2023self,gardiner1992wave,gardiner2004quantum,marcuzzi2016absorbing,
Buchhold2017Nonequilibriumeffectivefield}. 

In the following, we adopt a semiclassical continuum approach, replacing the operator formalism with the respective expectation values. 
The covariance of the classical Langevin noises derived from the underlying quantum Langevin noise operators is given by  $\expval{\xi_{l,t}^\nu \xi_{l',t'}^\beta} = \delta(t-t') \delta_{l,l'}\expval{\partial_t(\hat{\sigma}^\nu_{l,t}\hat{\sigma}^\beta_{l,t})-\partial_t(\hat{\sigma}^\nu_{l,t})\hat{\sigma}^\beta_{l,t}-\hat{\sigma}^\nu_{l,t}\partial_t(\hat{\sigma}^\beta_{l,t}) + \text{h.c.}}$ ($\nu,\beta\in\set{x,y,aa}$), where h.c. stands for the Hermitian conjugate~\cite{xiang2023self,gardiner1992wave,gardiner2004quantum,marcuzzi2016absorbing,
Buchhold2017Nonequilibriumeffectivefield}. 
We focus on the continuum limit, which is justified when the number of atoms within a facilitation sphere is sufficiently large. Under this condition, we define the active and total density fields as $\rho\left(\va{r},t\right) \equiv\Expval{ \text{Tr}\left\{\hat\sigma^{aa}_l\hat\rho \right\}}_{\va{r}}$ and $n\left(\va{r},t\right) \equiv \Expval{\text{Tr}\left\{\hat n_l\hat\rho \right\}}_{\va{r}}$, respectively. Here, $\expval{...}_{\va{r}}$ designates a spatial average over a facilitation sphere centered at position $\va{r}$~\cite{xiang2023self}.
The coherence fields, $\sigma^{x/y}$, are defined similarly. 

Based on the resultant Langevin equations for these classical fields, we further perturbatively eliminate the fast coherence fields $\sigma^{x/y}$ via the Janssen-De Dominicis-Martin-Siggia-Rose procedure~\cite{Janssen1976Lagrangeanclassicalfield, Martin1973StatisticalDynamicsClassical,onsager1953fluctuation,dominicis1976techniques}. This procedure, valid under quantum dephasing and weak driving, yields a path integral for the active density field~\cite{xiang2023self},
\begin{equation}
    \mathcal{Z}=\int\mathcal{D}[\rho,\tilde\rho]e^{-\mathcal{S}[\rho,\tilde\rho]}
\end{equation}
with the action
\begin{equation}
\label{eq:action}
    \mathcal{S}[\rho,\tilde\rho]=\int \tilde\rho\left[\left(\partial_t-D_\rho\nabla^2+u_2\right)\rho+u_3\rho^2+u_4\rho^3-\mu\tilde\rho\right]
\end{equation}
and the coupled Langevin equations for active and total densities~\cite{xiang2023self},
\begin{subequations}
\begin{align}
    &\partial_t\rho=D_{\rho}\nabla^2\rho-u_2\rho-u_3\rho^2-u_4\rho^3+\tau n+\xi_\rho\label{eq:drhodt}\\
    &\partial_t n=D_\text{T}\nabla^2 n -b\rho+\lambda+\xi_n\label{eq:dndt}
\end{align}
\end{subequations}
Here, the coefficients are given by $u_2 = 1 - n\kappa - 256 n^2\Omega^4/\left(n\kappa + \gamma\right)^7$, $u_3 = 2\left[\kappa - 2n\Omega^2/(n\kappa + \gamma)\right]$, and $u_4 = 8\Omega^2/(n\kappa + \gamma)$.
The terms $\xi_\rho$ and  $\xi_n$ represent Gaussian white noises with zero mean, characterized by the variance-covariance matrix elements $M_{\rho\rho} = \left(1 + n\kappa\right)\rho + 4n\Omega^2 \rho^2/\left(n\kappa + \gamma\right)^2=2\mu$ and $M_{nn} = b\rho$. Furthermore, the effective diffusion constant is defined as $D_\rho = D_\text{T} + n\kappa R_\text{fac}^2 / 2$, where $D_\text{T}$ accounts for the thermal diffusivity.

\section{Results}

\subsection{Effective local potential for active density}
With the total density $n$ held fixed, the active-density equation~\eqref{eq:drhodt} defines an effective local potential $\Phi(\rho)$ conditioned on $n$. Omitting spatial fluctuations, variation with respect to $\rho$ and $\tilde\rho$ yields the Hamilton-Jacobi equations~\cite{Kamenev2023Fieldtheorynon}
\begin{subequations}
\label{eq:emo_opa}
    \begin{align}
        &\partial_t \tilde\rho=-\partial_\rho H_\text{cl} =\tilde\rho\left(u_2+2u_3\rho+3u_4\rho^2\right) - \tilde\rho^2\left(\mu_1+2\mu_2\rho\right)\\
        &\partial_t \rho=\partial_{\tilde\rho} H_\text{cl}=-\left(u_2\rho+u_3\rho^2+u_4\rho^4\right)+2\mu\tilde\rho
    \end{align}
\end{subequations}
Here, the density-dependent coefficient is given by $\mu = \mu_1 \rho + \mu_2 \rho^2$, with $\mu_1 = (1 + n\kappa)/2$ and $\mu_2 = 2n\Omega^2 / (n\kappa + \gamma)^2$. The corresponding classical Hamiltonian, $H_{\text{cl}}$, is expressed as:
\begin{equation}
    H_\text{cl}=-\tilde\rho\left(u_2\rho+u_3\rho^2+u_4\rho^3\right)+\mu\tilde\rho^2
\end{equation}
The increment of the action along the optimal path (with $H_\text{cl}=0$) described in Eqs.~\eqref{eq:emo_opa} connecting a pair of noiseless ($\tilde\rho=0$) states is defined as the difference in the effective local potential between them~\cite{zakine2023minimum}.

By defining the absorbing state (with $\rho=0$) as the reference point where the potential vanishes [$\Phi(\rho=0)=0$], the effective local potential for an arbitrary state is given by,
\begin{equation}
    \begin{aligned}
        \Phi(\rho)=&\int_0^\rho dx\left\{\frac{u_2x+u_3x^2+u_4x^3}{\mu_1 x+\mu_2x^2} \right\} \\
        = &\frac{u_4}{\mu_2}\left[\frac{\rho^2}{2}+\left(\frac{u_3}{u_4}-\frac{\mu_1}{\mu_2}\right)\rho \right]\\
        &+ \frac{u_4}{\mu_2}\left[\frac{u_2}{u_4} + \frac{\mu_1}{\mu_2}\left(\frac{\mu_1}{\mu_2}-\frac{u_3}{u_4}\right)\right]\ln\left(\mu_2\rho+\mu_1\right)
    \end{aligned}
\end{equation}
Typical effective local potential curves for low and high total densities are shown in Fig.~\ref{fig:model}(b). 
At low total density $n$, the system favors the absorbing state, whereas high $n$ shifts the equilibrium toward the active state. The interplay between slow loss (at rate $b\Gamma$) and pumping (at rate $\lambda$) couples the fast transition dynamics to the slow evolution of the total density, triggering a LC phase along the hysteresis loop of a first-order APT, as described in detail below.

\subsection{First-order absorbing-state phase transition and tipping point}

\begin{figure}[t]
	\centering
	\includegraphics[width =8.6cm, keepaspectratio]{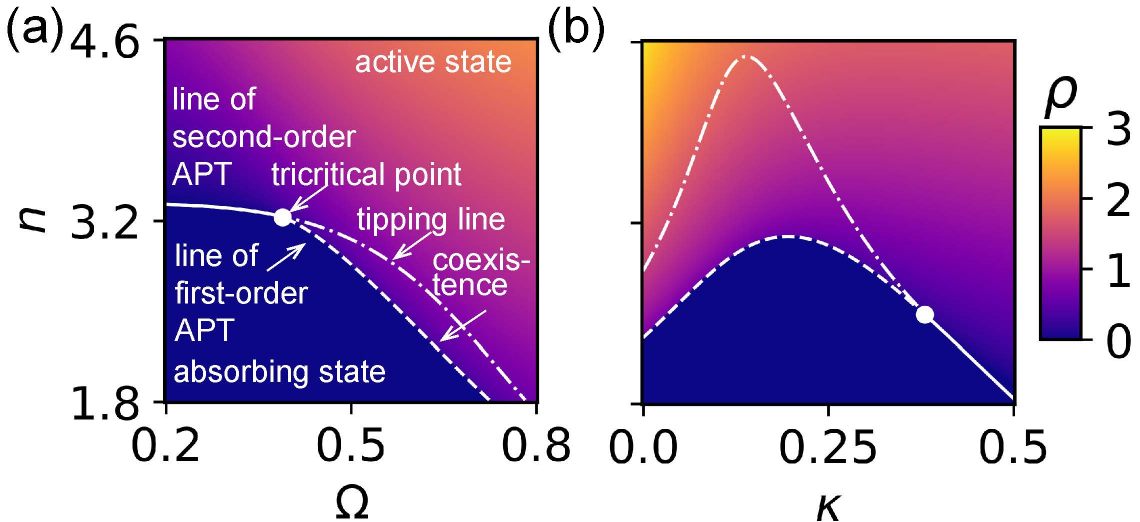}
	\caption{Mean-field (MF) phase diagrams for (a) $\kappa=0.3$ and (b) $\Omega=0.5$ at fixed total density $n$. Color indicates the active density $\rho$. In the classical (quantum) regime, the system exhibits a second-order (first-order) APT from the absorbing with $\rho=0$ to an active state with finite $\rho$, marked by solid (dashed) lines. The two transitions meet at a tricritical point (dot). 
    Within the first-order APT regime, the tipping point (dash-dotted lines) marks where the absorbing state becomes linearly unstable ($u_2=0$). The region between the tipping line and the first-order APT line is a coexistence regime where the absorbing and active states are both stable.}
	\label{fig:PD}
\end{figure}

Based on the effective action~\eqref{eq:action} for the active density, we obtain the MF phases by identifying the noiseless saddle-point solutions at a fixed total density $n$. The corresponding phase boundaries are illustrated in Fig.~\ref{fig:PD}.
In the quantum regime ($\Omega>\kappa$), the excitation density changes discontinuously from zero to a finite value as the total density $n$ crosses a critical threshold (dash-dotted line), marking a first-order APT. This discontinuous jump forms a key ingredient of the SOB-induced LCs discussed later. As the classical activation rate $\kappa$ increases, the system enters a classical regime characterized by a second-order transition (solid line), consistent with self-organized criticality in driven-dissipative Rydberg gases~\cite{klocke2019control,helmrich2020signatures, Ding_Phase_2020}.

In the coexistence regime, further increase of the total density destabilizes the absorbing state. This behavior is captured by the linear coefficient $u_2$, which characterizes the inverse recovery rate of a homogeneous system near the absorbing state.
When $u_2>0$, the absorbing state is linearly stable and fluctuations in the active density fields dampen over time. In contrast, $u_2<0$ indicates linear instability, where even an infinitesimal perturbation can trigger an abrupt surge in the active density.
The susceptibility of the system is also manifested in the spatial propagation of an active seed. To quantify this, we analyze the spatiotemporal fluctuations of the active density field at a fixed total density $n$. Linearizing Eq.~\eqref{eq:drhodt} near the absorbing state $\rho_{0,t}=0$ and performing a Fourier transform yields the following evolution equation,
\begin{equation}
\partial_t \delta\rho_{\va{k},t} = -(k^2 + u_2)\delta\rho_{\va{k},t} + \xi_\rho
\end{equation} 

Consider an initial configuration where all sites are absorbing except for a single seed localized at
$\va{r}_0$, expressed as $\rho_{\va{r},t=0}=a_0\delta\left(\va{r}-\va{r}_0\right)$, with $0<a_0\ll 1$. Applying a Laplace transform leads to
\begin{equation}
\label{eq:susceptibility_k}
(k^2+u_2+i\omega)\delta\rho_{\bar\omega} = \xi_\rho + a_0 e^{-i\va{k}\cdot\va{r}_0}
\end{equation} 
where the shorthand notation $\bar\omega=(\omega,\boldsymbol{k})$. After averaging over the noise and performing the inverse Fourier transform, the real-space dynamics are given by,
\begin{equation}
\label{eq:susceptibility_r}
\frac{\Expval{ \delta\rho_{\va{r},t} } }{a_0}=\int{d\bar\omega \frac{ e^{i\left[\omega t + \boldsymbol{k}\cdot(\va{r}-\va{r}_0)\right]}}{k^2+u_2+i\omega}}\propto \frac{e^{-\frac{\Abs{ \va{r}-\va{r}_0}^2}{4t} - u_2 t}}{t^{d/2}}
\end{equation} 
where $d$ denotes the spatial dimension. Consequently, for $u_2>0$, the activation remains exponentially localized within a characteristic length scale of $\xi \sim u_2^{-1/2}$. In contrast, for $u_2<0$, the initial seed proliferates, leading to a spreading activation front across the entire system.

Consequently, within the first-order APT regime, any parameter $x\in\{n,\Omega,\kappa\}$ satisfying  $u_2=0$ identifies a \emph{tipping point} \cite{wissel1984a,lenton2008tipping,veraart2012recovery,van2007slow}, marking the onset of propagating activation seeds [represented by dash-dotted lines in Fig.~\ref{fig:PD}]. Our analysis focuses on this threshold while disregarding the secondary transition from the active to the absorbing state [the dashed lines in Fig. 2];
Unlike the correlated collective activation that drives the tipping point, the latter is governed merely by stochastic, local, and uncorrelated inactivation events.

\begin{figure}[t]
    \centering
	\includegraphics[width = 8.4 cm, keepaspectratio]{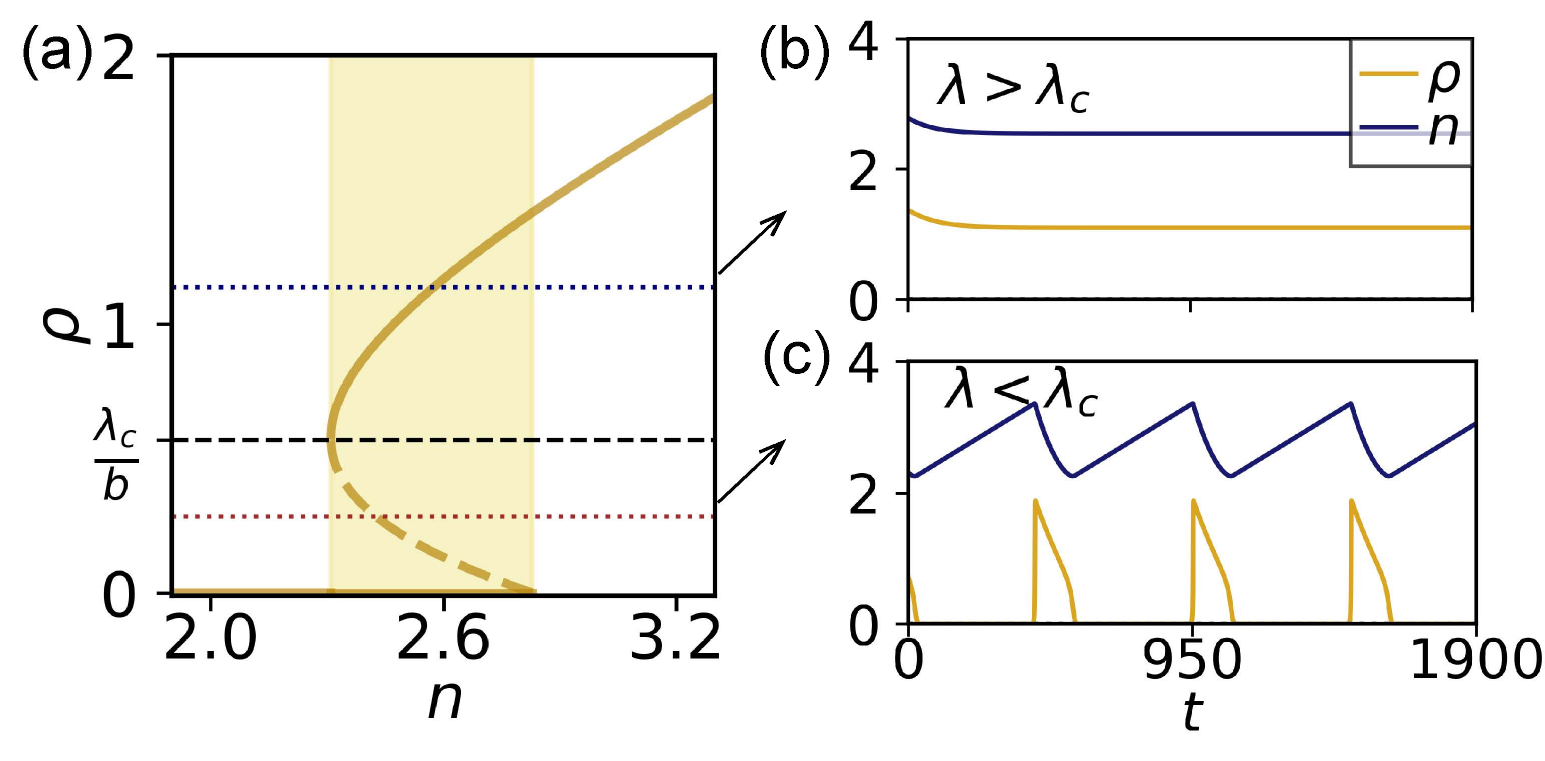}
	\caption{MF picture of limit cycle state induced by self-organized bistability. (a) For a given loading rate $\lambda$, the stationary solution to the coupled equations~\eqref{eq:drhodt} and~\eqref{eq:dndt} corresponds to the intersection between the yellow curve (fixed points of the active-density equation~\eqref{eq:dndt} at fixed total density; solid/dashed segments indicate stable/unstable branches) and the horizontal dashed line (fixed point of total-density equation~\eqref{eq:dndt}).
    The coexistence regime is shaded yellow.
    A critical loading rate $\lambda_c$ separate two regimes: (b) limit-cycle state for $\lambda<\lambda_c$ and (c) stationary state for $\lambda>\lambda_c$.}
	\label{fig:SOB}
\end{figure}

\subsection{The realization of limit cycle state}
In the regime of first-order APTs, introducing loading and loss processes couples the dynamics of the active density to its control parameter, the total density. Within the coexistence regime, as illustrated in Fig.~\ref{fig:SOB}(a), the fixed point of the equation for total density~\eqref{eq:dndt} can be either stable or unstable fixed point of the equation for active density~\eqref{eq:drhodt}, depending on the loading rate $\lambda$.
A critical loading rate, $\lambda_c$, is defined as the condition where the steady-state active density that nullifies Eq.~\eqref{eq:dndt} coincides with the active density at the critical total density of the first-order phase transition (represented by the black dashed line).
This critical loading rate demarcates two distinct regimes: a stationary state for $\lambda > \lambda_c$ [Fig.~\ref{fig:SOB}(b)] and a LC regime for $\lambda < \lambda_c$ [Fig.~\ref{fig:SOB}(c)].

In Eq.~(\ref{eq:drhodt}), a weak, rare spontaneous activation process at a rate $\tau \ll \kappa,\Omega$ prevents the system from being permanently trapped in the absorbing state.
Without this term, the activation dynamics would cease, resulting in overloading.
Once the absorbing state is reached, this rare process restores the dynamics over a characteristic timescale of $(\tau N)^{-1}$,  where $N$ is the total number of emitters. This mechanism counteracts potential overloading and is essential for sustaining long-term phase coherence in finite-sized systems~\cite{xiang2023self}.

Furthermore, in finite dimensions, the robustness of these LCs is intimately linked to the underlying first-order transition; consequently, they are expected to persist for $d\geq2$ \cite{xiang2023self}. For simplicity, the following analysis focuses on the three-dimensional case where stable LCs have been established, using the parameter set $D_\text{T},D_\rho=1,\kappa=0,\Omega=0.5,\gamma=2,\tau=10^{-7}$, and $b=10^{-2}$.

\subsection{Linear stability of limit cycle state}

\begin{figure}[t]
    \centering
	\includegraphics[width = 8.4 cm, keepaspectratio]{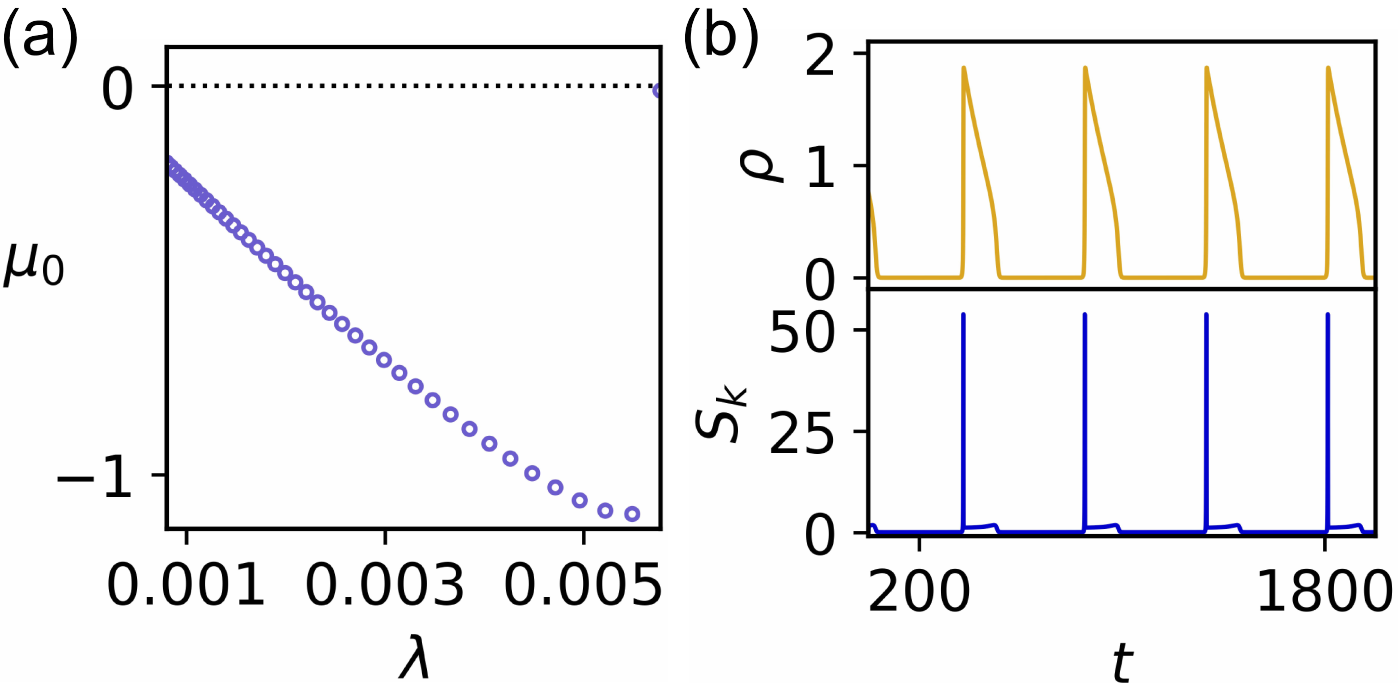}
	\caption{(a) Floquet multiplier that corresponds to fluctuations of active density fields with respect to MF trajectory as a function of $\lambda$ for $k=0$. (b) The dynamical behavior of the active density (upper) and correlation function (lower) $S_k=\left\langle \delta\rho_{\boldsymbol{k},t} \delta\rho_{-\boldsymbol{k},t}\right\rangle$ within Gaussian approximation for $k=0.1\pi$. In line with a negative Floquet multiplier in (a), the correlation function $S_k$ is periodic and finite with peaks corresponding to the tipping point where the system is about to jump from the absorbing state towards the absorbing state and is therefore susceptible to fluctuations.}
	\label{fig:floquet}
\end{figure}

We now demonstrate that the tipping point triggers instantaneous long-range correlations without compromising the stability of the LC. To investigate this, we consider the fluctuations $\delta\rho_{\va{r},t}$ and $\delta n_{\va{r},t}$  around the MF trajectory ($\rho_{0,t}$, $n_{0,t}$) within a Gaussian approximation. By expanding Eqs.~\eqref{eq:drhodt} and~\eqref{eq:dndt} and performing a Fourier transform, we define the fluctuation fields as $\delta\rho_{\boldsymbol{k},t}=\rho_{\boldsymbol{k},t}-\rho_{0,t}$ and $\delta n_{\boldsymbol{k},t} =n_{\boldsymbol{k},t}- n_{0,t}$. 
The linearized equations of motion for the correlation functions, written in the vector form 
$\boldsymbol{C}_{k,t}=
\begin{pmatrix}
\Expval{ \delta\rho_{\va{k},t}\delta\rho_{-\va{k},t} }, & \Expval{\delta\rho_{\va{k},t} \delta n_{-\va{k},t} }, & \Expval{ \delta n_{\va{k},t}\delta n_{-\va{k},t} }
\end{pmatrix}^T$, are governed by,
\begin{equation}
\label{eq:dCdt}
\partial_t \va{C}_{k,t}= \va{L}_{k,t} \va{C}_{k,t} + 2 \va{M}_{k,t}
\end{equation}
where the superscript `$T$' denotes the transpose, and the noise covariance vector is $\boldsymbol{M} =
\begin{pmatrix}
M_{\rho\rho}, &0,&M_{nn}
\end{pmatrix}^T
$. The linear response matrix $\boldsymbol{L}(k,t)$ is derived from the Jacobian matrix $J_{\alpha\beta} = \left(\partial_\beta F_\alpha \right)_0 - \delta_{\alpha,\beta}k^2$ (with $\alpha,\beta\in\set{\rho,n}$),  where $F_\rho$ and $F_n$ represent the deterministic components of Eqs.~\eqref{eq:drhodt} and~\eqref{eq:dndt}, respectively. The subscript `0' indicates evaluation along the MF values $\rho=\rho_{0,t}$ and $n=n_{0,t}$. 
The matrix $\boldsymbol{L}$ is then explicitly given by,
\begin{equation}
\label{eq:drift}
 \boldsymbol{L} = 
\begin{pmatrix}
2J_{\rho\rho} & 2J_{\rho n}  & 0 \\
J_{n \rho} & J_{\rho\rho}  + J_{nn} &J_{\rho n} \\
 0 &2 J_{n\rho} & 2J_{nn} 
\end{pmatrix}
\end{equation}
Due to the time periodicity of $\rho_{0,t}$ and $n_{0,t}$, the drift matrix $\boldsymbol{L}$ and the diffusion vector $\boldsymbol{M}$ are inherently time-dependent. Consequently, the correlation functions $\va{C}_{k,t}$ also exhibit corresponding temporal oscillations.
Given that the dynamics of the active and total density fields occur on widely separated timescales, we can employ a separation of variables approach. By neglecting all correlation components in Eq.~\eqref{eq:dCdt} except for the active-density autocorrelation $S_{k,t}=\left\langle \delta\rho_{\boldsymbol{k},t}  \delta\rho_{-\boldsymbol{k},t}\right\rangle$, the evolution simplifies to,
\begin{equation}
\label{eq:dSkdt}
\partial_t S_{k,t} = 2J_{\rho\rho} S_{k,t} + 2M_{\rho\rho}
\end{equation}
For the $m$-th period, defined as $t\in\left((m-1)T,mT\right]$ with $T$ being the limit cycle period, the formal solution under the initial condition $S_{k,0} = 0$ is given by:
\begin{equation}
\label{eq:S_k_formal}
\begin{aligned}
S_{k,t} &= 2\int_0^t dx M_{\rho\rho}(x) e^{2\int_x^t{dy J_{\rho\rho}(y)}}\\
& = 2\left(\sum_{l=0}^{l=m-1}\int_{lT}^{(l+1)T}  + \int_{mT}^t\right)dx\left\{M_{\rho\rho} e^{2\int_x^t{dy J_{\rho\rho}}}\right\}\\
& = S_{k,t-mT} + 2e^{\int_{0}^{t-mT}{dx J_{\rho\rho}(x)}} S_{k,T} \frac{1-e^{mT\mu_k}}{1-e^{T\mu_k}}
\end{aligned}
\end{equation}
where the Floquet multiplier~\cite{boland2009limit,boland2008how, geysermans1997stochastic,chan2015limit,strogatz2018nonlinear} is given by
\begin{equation}
\label{eq:floquet_multiplier}
\mu_k = \frac{2}{T}\int_0^T{dx J_{\rho\rho}(x)} = \mu_0 - 2k^2
\end{equation}
It follows from Eq.~\eqref{eq:S_k_formal} that the LC becomes unstable if $\mu_k \geq 0$ for any $k$.

The zero-mode Floquet multiplier $\mu_0$ (at $k=0$), obtained via numerical integration of Eq.~\eqref{eq:floquet_multiplier}, is presented in Fig.~\ref{fig:floquet}(a). Notably, $\mu_0$ remains negative throughout the parameter regime where the LC exists. Since $\mu_k < \mu_0$ for all $k > 0$, all finite-$k$ modes are likewise stable. Consequently, the correlation function $S_{k,t}$ remains bounded, consistent with the results from the direct numerical integration of Eq.~\eqref{eq:dSkdt} [see Fig.~\ref{fig:floquet}(b)].

As illustrated in Fig.~\ref{fig:floquet}(b), $S_k$ manifests a sharp peak within each period, signaling a transient but pronounced enhancement in the system's response to perturbations (lower panel) immediately prior to the transition from the absorbing to the active state (upper panel). 
As discussed in the previous section, this enhanced sensitivity to fluctuations coincides with the linear instability of the absorbing phase, which occurs when the total density $n$ satisfies $J_{\rho\rho}|_{\rho=0} = u_2(n) = 0$. This condition corresponds precisely to the tipping point indicated by the dash-dotted lines in Fig.~\ref{fig:PD}.
At this point, the system reacts strongly to fluctuations in the active field, triggering an abrupt surge in the active density. 
Importantly, this phenomenon does not signify an instability of the LC itself; rather, it reflects a dramatic spike in dynamical susceptibility during the switching from the absorbing to the active state.

In the following, we present the results obtained from numerical simulations of Eqs.~\eqref{eq:drhodt} and~\eqref{eq:dndt}. These simulations employ an operator-splitting scheme~\cite{pechenik1999interfacial,dornic2005integration,xiang2023self} to ensure numerical stability and accuracy.

\subsection{State-dependent entropy production rate}
\begin{figure}[t]
     \centering
	\includegraphics[width = 6.5 cm, keepaspectratio]{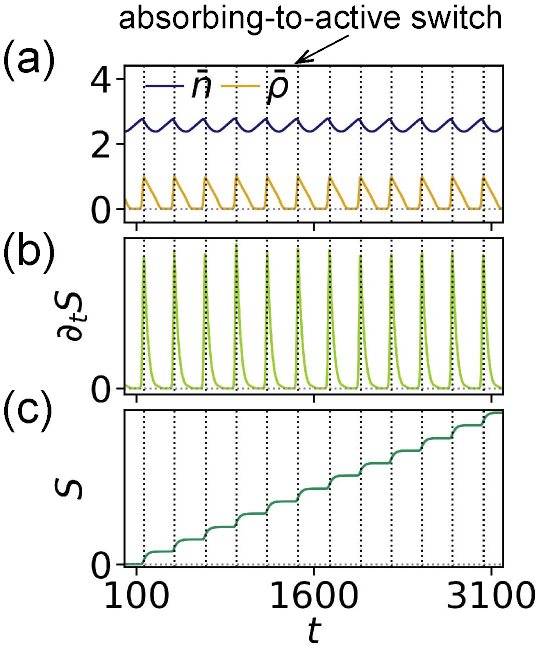}
	\caption{Time series of (a) the spatially averaged total $\bar n$ and active $\bar\rho$ densities, (b) instantaneous entropy production rate $\partial_t S$, and (c) total entropy production $S$. Gray dashed lines mark transitions from the absorbing to the active state. The instantaneous entropy production rate peaks at the tipping point where the system switches from the absorbing to active 
    states. Parameters for (a)-(c) are $L=64,\lambda=3.2\times10^{-3}$. }
	\label{fig:entro}
\end{figure}

LCs are nonequilibrium species that break time translation invariance. 
To sustain such dynamical asymmetry, these systems require a continuous influx of energy and a persistent rate of entropy production~\cite{ao2004potential, wang2008potential}. In this context, entropy associated with the breaking of time-reversal symmetry is defined as dynamical entropy, distinguishing it from the static entropy related to spatial symmetries. This dynamical entropy provides a rigorous measure of path irreversibility, typically expressed as the log-ratio between the probabilities of forward and time-reversed trajectories~\cite{gaspard2004time}.
Correspondingly, the steady-state entropy production rate per unit volume, arising from the dynamics of the active density field, is defined as
\begin{equation}
\label{eq:dtdS_def}
\partial_t S(t)=\frac{1}{L^d}\sum_{l}{ \int{ \frac{d\rho d\rho'}{dt} p_{l,t}(\rho,\rho')\ln\left[\frac{p_{l,t}(\rho,\rho')}{p_{l,t}^\text{R}(\rho',\rho)}\right]}  }
\end{equation}
In this expression, $p_{l,t}(\rho, \rho')$ represents the probability for a local trajectory at site $l$ to evolve from $\rho$ to $\rho'$ over a short interval $dt$. The superscript “$\text{R}$” specifies the time-reversed trajectory. 

In the steady state, spatial translation invariance allows the marginal probability to be expressed as $p_{l,t}(\rho, \rho') = \int dn P^{\text{SS}}_t(n, \rho) P(\rho \to \rho')$, where $P^{\text{SS}}_t(n, \rho)$ is the steady-state probability density of average total and active densities at time $t$, and $P(\rho \to \rho')$ is the short-time propagator associated with the Langevin equation~\eqref{eq:drhodt}. Within the Onsager-Machlup framework~\cite{Janssen1976Lagrangeanclassicalfield, Martin1973StatisticalDynamicsClassical, onsager1953fluctuation, dominicis1976techniques}, this transition probability takes a Gaussian form:
\begin{equation}
P(\rho \to \rho') = \frac{1}{\sqrt{2\pi M_{\rho\rho}dt}} \exp\left[ -\frac{(\rho' - \rho - F_{\rho}dt)^2}{2M_{\rho\rho}dt} \right]
\end{equation}
where the deterministic drift $F_{\rho}$ is evaluated at the initial state $\rho$. Its time-reversed counterpart $P^{\text{R}}(\rho' \to \rho)$ is obtained by exchanging the initial and final states while maintaining the functional form of the drift. The backward path probability is thus $p^{\text{R}}_{l,t}(\rho', \rho) = \int dn P^{\text{SS}}_t(n, \rho) P^{\text{R}}(\rho' \to \rho)$~\cite{gaspard2004time}. By substituting the stochastic evolution $\rho' = \rho + (F_\rho + \xi_\rho)dt$ into the definition of the entropy production rate~\eqref{eq:dtdS_def}, we arrive at the following integral expression:
\begin{equation}
\partial_t S(t)= \int { d\rho dn d\xi_\rho \frac{ e^{-\frac{\xi_\rho^2}{2M_{\rho\rho} } } }{\sqrt{2\pi M_{\rho\rho}}}P^\text{SS} _t(n,\rho) \frac{2F_\rho (F_\rho+\xi_\rho) }{M_{\rho\rho}}}
\end{equation}
After integrating over the noise $\xi_\rho$, the entropy production rate simplifies to the following  form,
\begin{equation}
\partial_t S(t) = 2 \left\langle \frac{F_\rho^2}{M_{\rho\rho}} \right\rangle_l
\end{equation}
where $\langle \dots \rangle_l$ denotes the spatial average over all sites. 
This expression corresponds to the generalized energy dissipation rate for the overdamped Langevin dynamics in Eq.~\eqref{eq:drhodt}. Specifically, it represents the ratio of the squared generalized force to its associated diffusion coefficient~\cite{doi2013soft, onsager1953fluctuation, bertini2015mft}. Consequently, this result highlights the fundamental balance between entropy production and energy dissipation within the system's non-equilibrium steady state.

Temporal integration of the entropy production rate, $\partial_t S$, yields the total accumulated entropy, $S$. As illustrated in Fig.~\ref{fig:entro}(b), the entropy production rate exhibits a sharp peak during the transition from the absorbing to the active state, which manifests as a characteristic step-like rise in the accumulated entropy $S$ [Fig.~\ref{fig:entro}(c)]. In regions away from this transition, the entropy production increases only marginally and remains nearly stationary within the absorbing state.
This intermittent burst of entropy production serves as a thermodynamic signature of the discontinuous switching process, reflecting the dissipative cost required to cross the tipping point manifold.

\subsection{Dynamical spatial correlations}

\begin{figure}[t]
     \centering
	\includegraphics[width = 7 cm, keepaspectratio]{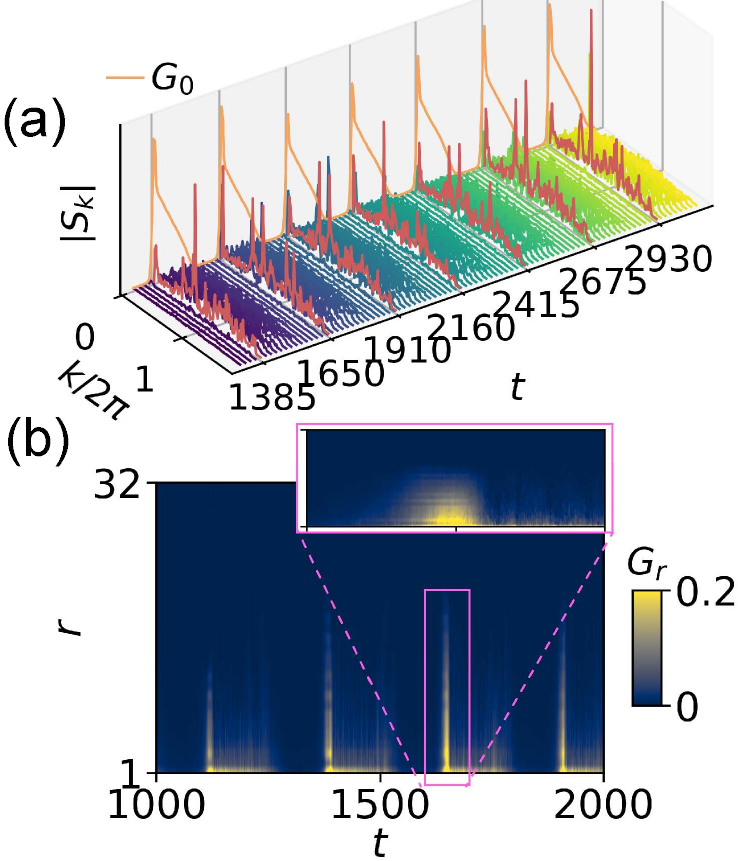}
	\caption{Dynamical spatial correlations of limit cycle state. (a) Amplitude of the pairwise correlation functions of the active density field in Fourier space, $\left\vert S_k \right\vert$, colored by time.
    The curve is highlighted in red when the absorbing-to-active transition occurs (also marked by gray lines therein). The auto-correlations $G_0$ are colored in orange. 
    (b) Temporal behavior of real-space correlation functions $G_r$ (color-coded) where the insert provides a zoomed-in view of how long range correlations emerge and disappear as the system crosses the tipping point enclosed by the magenta frame. 
    Parameters for (a) and (b) are $L=64,\lambda=3.2\times10^{-3}$.}
	\label{fig:sim}
\end{figure}

Building upon the analysis of global correlations and entropy production, we now examine the spatial organization of the system. This is quantitatively assessed through two-point spatial correlation functions, which reveal the underlying structural order and collective behavior. Due to the system's invariance under rotation and displacement, the real-space correlation function $G_{r,t}$ depends solely on the relative distance $r$,
\begin{equation}
G_{r,t}=\Expval{ \delta\rho_{l,t} \delta\rho_{k,t} \delta\left(r-\left\vert \boldsymbol{r}_l - \boldsymbol{r}_k\right\vert\right) }_{l,k}
\end{equation}
whose Fourier transform yields the static structure factor $S_{k,t}$ \cite{doi2013soft}. Under the assumption that non-local correlations are subordinate to those induced by quantum noise, the steady-state auto-correlation can be estimated as $G_0 \approx \bar{\rho}/2d$.

Transitions from the absorbing to the active state are accompanied by sharply enhanced spatial correlations, as illustrated in Figs.~\ref{fig:sim}(a) and (b). These results align with our Gaussian analysis, where the dynamical instability near the tipping point—manifesting as a sudden surge in active density—triggers a transiently heightened susceptibility to perturbations [Fig.~\ref{fig:floquet}(b)]. As the system approaches and then passes the tipping point, the rapid proliferation of active emitters and the concomitant emergence of long-range correlations resemble a wildfire-like spread through a dense emitter network. This burst of collective activation is subsequently quenched by the depletion of the emitter population, leading to the eventual disappearance of long-range order.

\subsection{Experimental Implementation}
\begin{figure}[b]
        \centering
        \includegraphics[width=\linewidth]{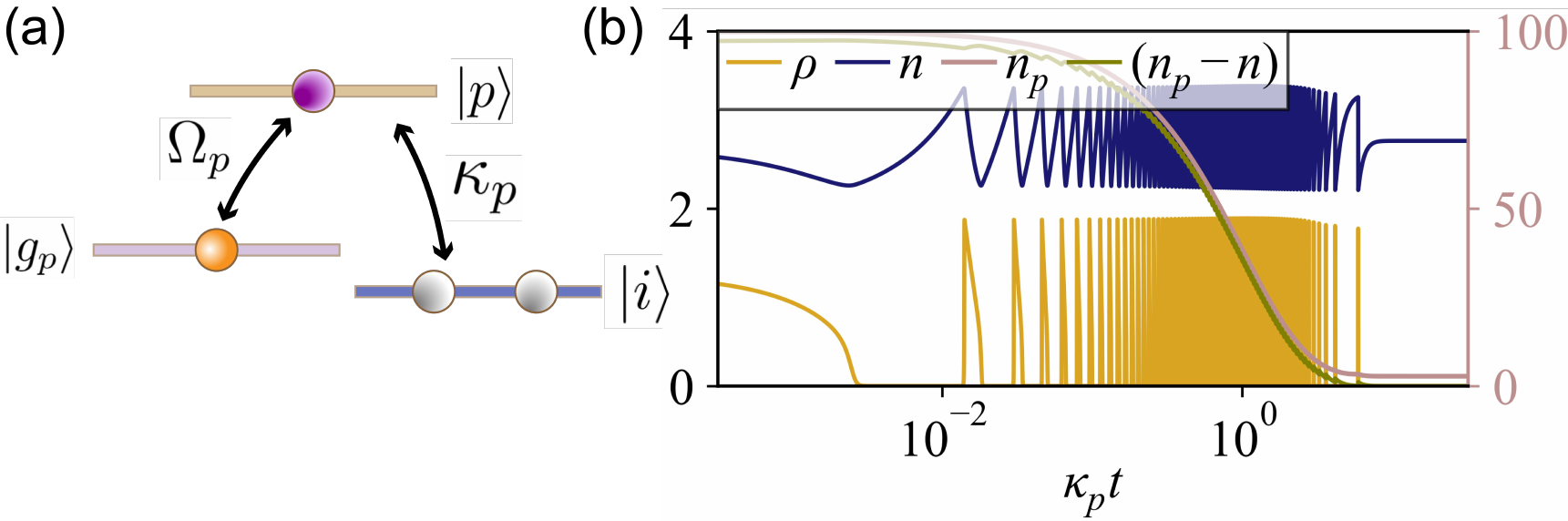}
        \caption{
        (a) Level scheme for pump process. Emitters are optically driven from the auxiliary ground state $\ket{g_p}$ to the pump state $\ket{p}$, which then transfers population to the inactive state $\ket{i}$ via an incoherent coupling with strength $\kappa_p$.
        (b) MF time series of active ($\rho$), total ($n$), and pump ($n_p$) particle densities for the quantum limit cycle. The left axis shows $\rho$ and $n$; the right axis shows $n_p$ and its difference from total density $(n_p - n)$. Time is rescaled by the pumping rate (incoherence coupling rate) $\kappa_p = \lambda/n_p(t=0)$, with initial reservoir density $n_p(0)=100$ and injection parameter $\lambda=3.2\times10^{-3}$.}
        \label{fig:pump}
    \end{figure}

Finally, we discuss potential experimental implementations. The proposed model can be realized using coherent laser-driven Rydberg atoms in the facilitation (anti-blockade) regime. In this setup, the electronic ground and Rydberg states are mapped to the inactive and active states, respectively, while additional auxiliary levels can represent the removed state~\cite{marcuzzi2016absorbing, klocke2019control, helmrich2020signatures, brady2024Anomalousb}. The facilitation radius in our model corresponds to the Rydberg blockade radius. Since our framework focuses on two-body contact processes, its validity holds for densities where at least two Rydberg atoms reside within a single facilitation sphere. 
Furthermore, the relative weight between coherent and incoherent activation can be tuned via the driving lasers.

Meanwhile, the injection of inactive emitters can be implemented via atomic pumping using Raman processes~\cite{wang2000Gainassisted, Deng2007gain}.
The pump state $\ket{p}$ is initially prepared by laser-coupling an auxiliary ground state $\ket{g_p}$ to $\ket{p}$ to achieve a population inversion, thereby establishing the initial population $n_p(t=0)$, and pump process is then realized through an incoherent coupling between $\ket{p}$ and $\ket{i}$ [see Fig.~\ref{fig:pump}(a)].
In a real experimental setup, however, the population in the pump state will eventually deplete over time.
If the system is initialized with a high population in the pump state and a low pumping rate $\kappa_p$, the limit cycle can last for many cycles.
This is demonstrated in Fig.~\ref{fig:pump}(b), where we initialize the system with a high pump-state density $n_p(0)=100$ (corresponding to 100 emitters within a single facilitation sphere) and a low pump rate $\kappa_p = \lambda/n_p(0)$, with $\lambda=3.2 \times 10^{-3}$.
The limit cycle persists for many cycles, exhibiting small variations in oscillation amplitude alongside a steady decrease in the pump-state density, until the dynamics halts when $n_p=n$.

\section{Discussions and Conclusion}

To summarize, in this work, we demonstrate the emergence of quantum LCs that exhibit dynamical spatial correlations and entropy production in an open quantum system. The periodic long-range correlations originate from the multiplicative quantum noises inherent to the system and from the repeated sweeping through a tipping point. The features are captured qualitatively by a Gaussian-Floquet analysis and confirmed quantitatively by numerical simulations. 

However, it is important to note that the long-range correlations stemming from the tipping point are insufficient to stabilize sustained oscillations—stable LCs are essential. 
Rather, it is the combined effects of three key elements: (i) the enhanced dynamical susceptibility near the tipping point, (ii) the active seed generation through the rare spontaneous activation process, and (iii) the independent local dissipation, which together produces the regular, periodic, back-and-forth collective switching between the absorbing and the active state. As our previous analysis indicates, the robustness of SOB-induced LCs is directly linked to the stability of first-order APTs under quantum fluctuations~\cite{xiang2023self}.
Therefore, in low spatial dimensions where first-order APTs are absent, the corresponding lack of tipping points precludes for the emergence of such dynamical synchrony among distant emitters. 

By demonstrating how tipping points give rise to long-range correlations in strongly interacting driven-dissipative quantum many-body systems, our work extends the notion of critical thresholds and early-warning signals to the far-from-equilibrium quantum realm~\cite{ding_tipping_2024}. Moreover, it also shows how SOB—as a certain type of Hopf bifurcation~\cite{cross1993pattern,strogatz2018nonlinear}—can aid in stabilizing the phase coherence of LCs in diffusively coupled, spatially extended, and noisy systems through self-organized correlations.

%

%



\section*{acknowledgment}
This work was supported by the National Natural Science Foundation of China (12347102 and 12547183), the Natural Science Foundation of Jiangsu Province (BK20233001), the Innovation Program for Quantum Science and Technology (2024ZD0300101), and the Fundamental and Interdisciplinary Disciplines Breakthrough Plan of the Ministry of Education of China (JYB2025XDXM502). The authors also acknowledge the computational resources provided by the High Performance Computing Center of Nanjing University.

%

\end{document}